\documentclass[preprint]{aastex}

\shorttitle{Exploring the Morphology of RAVE Stellar Spectra}
\shortauthors{Matijevi\v c et al.}

\begin{document}
\title{Exploring the Morphology of RAVE Stellar Spectra}

\author{G.~Matijevi\v c}
\affil{University of Ljubljana, Faculty of Mathematics and Physics, Jadranska 19, 1000 Ljubljana, Slovenia}
\email{gal.matijevic@fmf.uni-lj.si}

\author{T.~Zwitter}
\affil{University of Ljubljana, Faculty of Mathematics and Physics, Jadranska 19, 1000 Ljubljana, Slovenia}
\affil{Center of Excellence SPACE-SI, A\v sker\v ceva cesta 12, 1000 Ljubljana, Slovenia}

\author{O.~Bienaym\' e}
\affil{Observatoire de Strasbourg, Universit\'e de Strasbourg, CNRS, 11 rue de l'universit\'e, 67000 Strasbourg, France}

\author{J.~Bland-Hawthorn}
\affil{Sydney Institute for Astronomy, University of Sydney, NSW 2006, Australia}

\author{C.~Boeche}
\affil{Astronomisches Rechen-Institut, Zentrum f\"ur Astronomie der Universit\"at Heidelberg, M\"onchhofstr.\ 12--14, 69120 Heidelberg, Germany}

\author{K.~C.~Freeman}
\affil{Research School of Astronomy and Astrophysics, Australia National University, Weston Creek, Canberra ACT 2611, Australia}

\author{B.~K.~Gibson}
\affil{University of Central Lancashire, Jeremiah Horrocks Institute, Preston, PR1 3TE, UK}

\author{G.~Gilmore}
\affil{Institute of Astronomy, University of Cambridge, Madingley Road,
Cambridge CB3 0HA, UK}
\affil{Astronomy Department, Faculty of Science, King Abdulaziz University, P.O. Box 80203, Jeddah 21589, Saudi Arabia}

\author{E.~K.~Grebel}
\affil{Astronomisches Rechen-Institut, Zentrum f\"ur Astronomie der Universit\"at Heidelberg, M\"onchhofstr.\ 12--14, 69120 Heidelberg, Germany}

\author{A.~Helmi}
\affil{Kapteyn Astronomical Institute, University of Groningen, P.O. Box 800, 9700 AV Groningen, the Netherlands}

\author{U.~Munari}
\affil{INAF Osservatorio Astronomico di Padova, 36012 Asiago, Italy}

\author{J.~Navarro}
\affil{Department of Physics and Astronomy, University of Victoria, Victora, BC V8P 5C2, Canada}

\author{Q.~A.~Parker}
\affil{Department of Physics and Astronomy, Macquarie University, NSW 2109, Australia}
\affil{Macquarie Research Centre for Astronomy, Astrophysics and Astrophotonics
(MQAAAstro)}
\affil{Australian Astronomical Observatory, PO Box 296, Epping, NSW 2121,
Australia}

\author{W.~Reid}
\affil{Department of Physics and Astronomy, Macquarie University, NSW 2109, Australia}
\affil{Macquarie Research Centre for Astronomy, Astrophysics and Astrophotonics
(MQAAAstro)}

\author{G.~Seabroke}
\affil{Mullard Space Science Laboratory, University College London, Holmbury St Mary, Dorking, RH5 6NT, UK}

\author{A.~Siebert}
\affil{Observatoire de Strasbourg, Universit\'e de Strasbourg, CNRS, 11 rue de l'universit\'e, 67000 Strasbourg, France}

\author{A.~Siviero}
\affil{Department of Astronomy, Padova University, Vicolo dell'Osservatorio 2, 
35122 Padova, Italy}
\affil{Leibniz-Institut f\"{u}r Astrophysik Potsdam (AIP), An der Sternwarte 16, 14482 Potsdam, Germany}

\author{M.~Steinmetz}
\affil{Leibniz-Institut f\"{u}r Astrophysik Potsdam (AIP), An der Sternwarte 16, 14482 Potsdam, Germany}

\author{F.G.~Watson}
\affil{Australian Astronomical Observatory, PO Box 296, Epping, NSW 2121,
Australia}

\author{M.~Williams}
\affil{Leibniz-Institut f\"{u}r Astrophysik Potsdam (AIP), An der Sternwarte 16, 14482 Potsdam, Germany}

\author{R.~F.~G.~Wyse}
\affil{John Hopkins University, 3400 N Charles Street, Baltimore, MD 21218, USA}

\begin{abstract}
The RAdial Velocity Experiment (RAVE) is a medium resolution $(R\sim 7500)$ spectroscopic survey of the Milky Way which already obtained over half a million stellar spectra. They present a randomly selected magnitude-limited sample, so it is important to use a reliable and automated classification scheme which identifies normal single stars and discovers different types of peculiar stars. To this end we present a morphological classification of $\sim 350,000$ RAVE survey stellar spectra using locally linear embedding, a dimensionality reduction method which enables representing the complex spectral morphology in a low dimensional projected space while still preserving the properties of the local neighborhoods of spectra. We find that the majority of all spectra in the database $(\sim 90-95\,\%)$ belong to normal single stars, but there is also a significant population of several types of peculiars. Among them the most populated groups are those of various types of spectroscopic binary and chromospherically active stars. Both of them include several thousands of spectra. Particularly the latter group offers significant further investigation opportunities since activity of stars is a known proxy of stellar ages. Applying the same classification procedure to the sample of normal single stars alone shows that the shape of the projected manifold in two dimensional space correlates with stellar temperature, surface gravity and metallicity.
\end{abstract}


\section{Introduction}

In the era of automated astronomical surveys the amount of data they produce is overwhelming. This holds especially true for large spectroscopic surveys like the 2dF Galaxy Redshift Survey \citep{2001MNRAS.328.1039C}, the 6dF Galaxy Survey \citep{2004MNRAS.355..747J}, the Radial Velocity Experiment \citep[RAVE,][]{2006AJ....132.1645S}, the Sloan Digital Sky Survey \citep[SDSS,][]{2009ApJS..182..543A}, the Gaia-ESO Survey, the Hermes-Galah project, and the highly anticipated \textit{Gaia} mission. A common approach in getting an overview of the observed sample of spectra and learn more about its morphological diversity is to \emph{classify} spectra according to some criterion (a classical example is the MKK classification scheme for single stars).
When dealing with a large spectral sample, an automated approach becomes a necessity.
Several different dimensionality reduction numerical methods became popular for such tasks. One of the first uses of artificial neural networks for stellar spectra classification was performed by \citet{1994ApJ...426..340G} and \citet{1994MNRAS.269...97V}, followed by many other authors. Another frequently used technique applied to the same problem is principal component analysis (PCA). It was used by \citet{1995AJ....110.1071C} for galaxy spectra classification and later by other authors for the classification of stellar spectra \citep[e.g.][]{1997AJ....113.1865I,1998MNRAS.298..361B,2010AJ....139.1261M}. Lately, another method related to PCA named locally linear embedding \citep[LLE,][]{2000Sci...290.2323R} was used by \citet{2009AJ....138.1365V} for classification of SDSS galaxy spectra and by \citet{2011AJ....142..203D} for classification of SDSS stellar spectra. The latter method seems particularly suitable for classification purposes since it is able to grasp the complex spectral morphologies and project the spectra onto a low dimensional space where the correlations between the spectra can be studied more easily.

RAVE is an ongoing radial velocity survey aiming at measuring radial velocities of up to $10^6$ stars in the southern sky. Based on observations with the UK Schmidt Telescope, the experiment employs the 6dF multi-fiber instrument which is capable of recording up to 150 stellar spectra simultaneously. The wavelength range of the observed spectra ranges from $\sim 8420\,\mathrm{\AA}$ to $\sim 8780\,\mathrm{\AA}$ with a typical resolving power of $R\sim 7500$ (similar wavelength but with somewhat larger resolving power was selected for the instrument aboard the \textit{Gaia} mission). The selection of the observed stars is magnitude limited $(9<I<12)$. The reduction and analysis pipeline is designed to provide radial velocities as well as atmospheric parameters of stars. The latter are calculated by finding the best matching fit to the observed spectrum from a library of synthetic spectra by \citet{2005A&A...442.1127M}. During the reduction process all spectra are shifted to the zero radial velocity system. The data are made publicly available through incremental data releases. The latest one was published by \citet{2011AJ....141..187S}. Till this date, a couple of papers already discussed some types of peculiar spectra found in the RAVE survey. \citet{2009A&A...503..511M} focused on luminous blue variables in the Large Magellanic Cloud and \citet{2010AJ....140..184M} analyzed spectra of double-lined spectroscopic binary candidates.

Currently, the RAVE pipeline is lacking a classification processing stage. While it is known that the majority of spectra observed by RAVE belong to non-peculiar single stars, a quick browse though the spectra of already observed stars reveals that there is a significant number of both peculiar and problematic spectra present in the database. Since the stellar parameters pipeline presumes that all spectra can be properly fit by a single star synthetic spectrum, neglecting the outliers can lead to wrong results in both radial velocities and atmospheric parameters. Searching for peculiar spectra based on the goodness of fit between the observed and the synthetic spectrum is very unreliable and previous classification attempt \citep{2010AJ....140..184M} is only efficient at identifying double-lined spectroscopic binary spectra but it fails to give a reliable classification of the whole sample. To correct for these deficiencies, this study presents a consistent morphological classification of the RAVE spectra with three main goals: to (1) provide a clean sample without any peculiar or problematic spectra so further studies based on the results of the RAVE survey can be more reliable, (2) identify any interesting peculiar spectra, and (3) highlight all problematic spectra that were corrupted either at the observation or reduction stage so they can be reprocessed if possible, or discarded. 

The structure of the paper is as follows. Section \ref{sec:lle} gives an overview of the inner workings of the LLE method.  Section \ref{sec:proc} explains the  procedure we used to classify RAVE spectra and finally the results of the classification are given in Section \ref{sec:results} followed by a summary in Section \ref{sec:sum}.

\section{Locally Linear Embedding}\label{sec:lle}

LLE is a general dimensionality reduction method introduced by \citet{2000Sci...290.2323R}. The key feature and an advantage over some other similar approaches is that this method preserves the relations between neighboring points in a lower-dimensional projection. Points that are close together (by some definition of distance) in a high dimensional space, remain close together in a low-dimensional projected space. This makes it easier to discover different hidden relations between data points. The method itself is relatively simple and can be outlined in three main steps. A detailed description and derivation can be found in \citet{2000Sci...290.2323R}, \citet{2002dRD}, \citet{2003SR} and \citet{2009AJ....138.1365V}, on which the following is based on.

When applying this method to the spectral data, each dimension in the initial $D$ dimensional space is represented by each wavelength bin at which the spectrum is sampled, so each spectrum can be considered a point in the $D$ dimensional space. In order for spectra to be comparable among each other, the sampling points must be equal for all spectra. 

Having denoted the vector of $N$ input spectra with $\mathbf{x}=(\mathbf{x}_1,...,\mathbf{x}_N)$, we first need to find the $k$ nearest neighbors of each of the members of $\mathbf{x}$, were $k\ll N$. Note that the distance between the data points is metric dependent. Throughout this work we used euclidean distances. For later use we will denote the vectors of indexes of the nearest neighbors with
\begin{equation}
\mathbf{k}^i= \left\{ \begin{array}{rl}
 j, &\parbox[t]{6.7cm}{ if $j$-th spectrum is among the nearest neighbors of the $i$-th spectrum} \\
 0, &\parbox[t]{6.7cm}{ otherwise}
       \end{array} \right.
\end{equation}

Following the first step, the local geometry of each data point is characterized by a linear combination of its neighbors. This step requires that the manifold on which the data points lie is sampled well enough that for all points from the data set the linear approximation is sufficiently accurate. The cost function related to the reconstruction error is written as
\begin{equation}\label{eq:cost_func}
\mathcal{E}(\mathbf{w})=\sum_i   \Bigl| \textbf{x}_i - \sum_j w_j^i\textbf{x}_j \Bigr|^2.
\end{equation}
The index $i$ passes through all spectra in the data set and the index $j$ through all the $k$ nearest neighbors of the $i$-th spectrum. The weights $w^i_{j}$ describe the contribution of the $j$-th neighborhood spectrum to the reconstruction of the $i$-th one. To find a set of weights $\mathbf{w}=(\mathbf{w}_1,...,\mathbf{w}_N)$ that will optimally reconstruct all data points, we need to minimize the cost function $\mathcal{E}$, enforcing the requirement that all weights contributing to the reconstruction of a single spectrum must add up to 1,
\begin{equation}\label{eq:sum_one}
\sum_j w_j^i=1.
\end{equation}
This is done by applying the Lagrangian multiplier $\lambda_i$ to  Equation~(\ref{eq:cost_func}),
\begin{eqnarray}\label{eq:i_cost_func}
\mathcal{E}^i(\mathbf{w})&=&\Bigl| \sum_j w_j^i(\textbf{x}_i-\textbf{x}_j) \Bigr|^2\to\\ &&\to\sum_{j=1}^k\sum_{l=1}^k w_{j}^iw_{l}^iC_{jl}^i-2\lambda_i\left(1-\sum_j w_j^i\right),
\end{eqnarray}
where $\mathcal{E}^ i(\mathbf{w})$ is the cost function corresponding to the reconstruction of the $i$-th point, and
\begin{equation}
C_{jl}^i=\left(\textbf{x}_i-\textbf{x}_j\right)^T \left(\textbf{x}_i-\textbf{x}_l\right)
\end{equation}
is the neighborhood correlation matrix. The expression in Equation~(\ref{eq:i_cost_func}) follows from Equation~(\ref{eq:cost_func}) when taking into account Equation~(\ref{eq:sum_one}). Optimal weights are found to be equal to
\begin{equation}
w_j^i=\frac{\sum_l (C_{jl}^i)^{-1}}{\sum_{m}\sum_{n}(C_{mn}^i)^{-1}},
\end{equation}
so the inverse of the correlation matrix needs to be calculated. Since this is a computationally intensive operation, a faster way of calculating the weights is to solve the linear system
\begin{equation}
\sum_j C_{jl}^iw_l^i=1,
\end{equation}
and to rescale the weights to satisfy Equation~(\ref{eq:sum_one}). If it happens that the correlation matrix is almost singular, a small fraction of the identity matrix can be added to it. \citet{2009AJ....138.1365V} suggested that a small fraction of the trace of the matrix $\mathbf{C}$ can be used so that
\begin{equation}
\mathbf{C}\to \mathbf{C}+r\mathbf{I},
\end{equation}
where they used the value of $r=10^{-3}\mathrm{Tr}(\mathbf{C})$.

In the last step of the method we need to find the $d<D$ dimensional space $Y$ onto which the data points can be optimally mapped using the same set of weights $\mathbf{w}$ as calculated before. The cost function related to the projection error is given by
\begin{equation}\label{eq:cost_func_y}
\tilde{\mathcal{E}}(Y)=\sum_i \Bigl| \textbf{y}_i - \sum_j w_{j}^i\textbf{y}_j \Bigr|^2,
\end{equation}
where $y_i$ is the vector to the $i$-th point in the $Y$ space. Stacking all weights $w_{j}^i$ in a single $N\times N$ sparse matrix $\mathbf{W}$ so that each row of the matrix corresponds to a single weight vector $\mathbf{w}_i$ and individual weights $w_j^ i$ from $\mathbf{w}_i$ are positioned according to the elements from the vectors $\mathbf{k}_i$,
enables us to rewrite Equation~(\ref{eq:cost_func_y}) as
\begin{equation}
\tilde{\mathcal{E}}(Y)=\sum_{j=1}^n\sum_{l=1}^n M_{ij}\mathbf{y}_i^T\mathbf{y}_j=\mathrm{Tr}(\mathbf{Y}\mathbf{M}\mathbf{Y}^T)
\end{equation}
where $\mathbf{M}=(\mathbf{I}-\mathbf{W})(\mathbf{I}-\mathbf{W})^T$. To avoid the trivial solution $\mathbf{Y}=0$, we can subject this function to the constraint
\begin{equation}
(\mathbf{Y}\mathbf{Y}^T)=\mathbf{I}.
\end{equation}
Similar as before, we can use Lagrangian multipliers and after setting the derivatives to zero we finish up with
\begin{equation}
(\mathbf{M}-\mathbf{\Lambda})\mathbf{Y}^T=0,
\end{equation}
where $\mathbf{\Lambda}$ denotes the diagonal matrix of all Lagrangian multipliers. Eigenvectors with the smallest eigenvalues are the ones we are looking for since they minimize the $\tilde{\mathcal{E}}(Y)$ and the eigenvector with the zero eigenvalue can be omitted since it only contributes to translation in space \citep{2000Sci...290.2323R,2002dRD}.

Because of the dependence of the projection space $Y$ on the weights $\mathbf{w}$ and consequently on the nearest neighbors of each point, the shape of the final projection depends on the input sample. This means that whenever new points are added to the sample, the whole process of projecting them has to be redone. The calculation of the projection can also be problematic if the input sample is large. To overcome these drawbacks, \citet{2009AJ....138.1365V} proposed a method in which we calculate the projection for only a smaller but still representative subset (base) and then add points to an already defined low dimensional space. To do that, we first need to find the $k$ nearest neighbors for each new point, but only among members of the base subset. The calculation of weights remains the same as described above, but the projection of points to the low dimensional space is computationally much cheaper since all that needs to be calculated are new vectors $\mathbf{y}_i$,
\begin{equation}
\mathbf{y}_i=\sum_j w_{j}^i\mathbf{y}_j,
\end{equation}
where $\mathbf{y_j}$ are vectors to the $k$ nearest neighbors in the $Y$ space corresponding to the $i$-th new point.

Several steps of the described method offer a possibility for a significant speed increase if some approximative approaches are used, as already noted by \citet{2009AJ....138.1365V}. In our calculations we used the FLANN library \citep{muja_flann_2009} for finding the $k$ nearest neighbors. The calculation of weights is an easily parallelizable task and can be spread among multiple processors with a linear speedup. Finally, eigenvectors of sparse matrices can be iteratively calculated with Arnoldi iterations which produce only a few sought-after eigenvectors instead calculating them all. In our case we used the routines from the ARPACK library. The code is available upon request from the authors.

\section{Classification Procedure}\label{sec:proc}

\begin{figure}
\begin{center}
\includegraphics[width=0.5\textwidth]{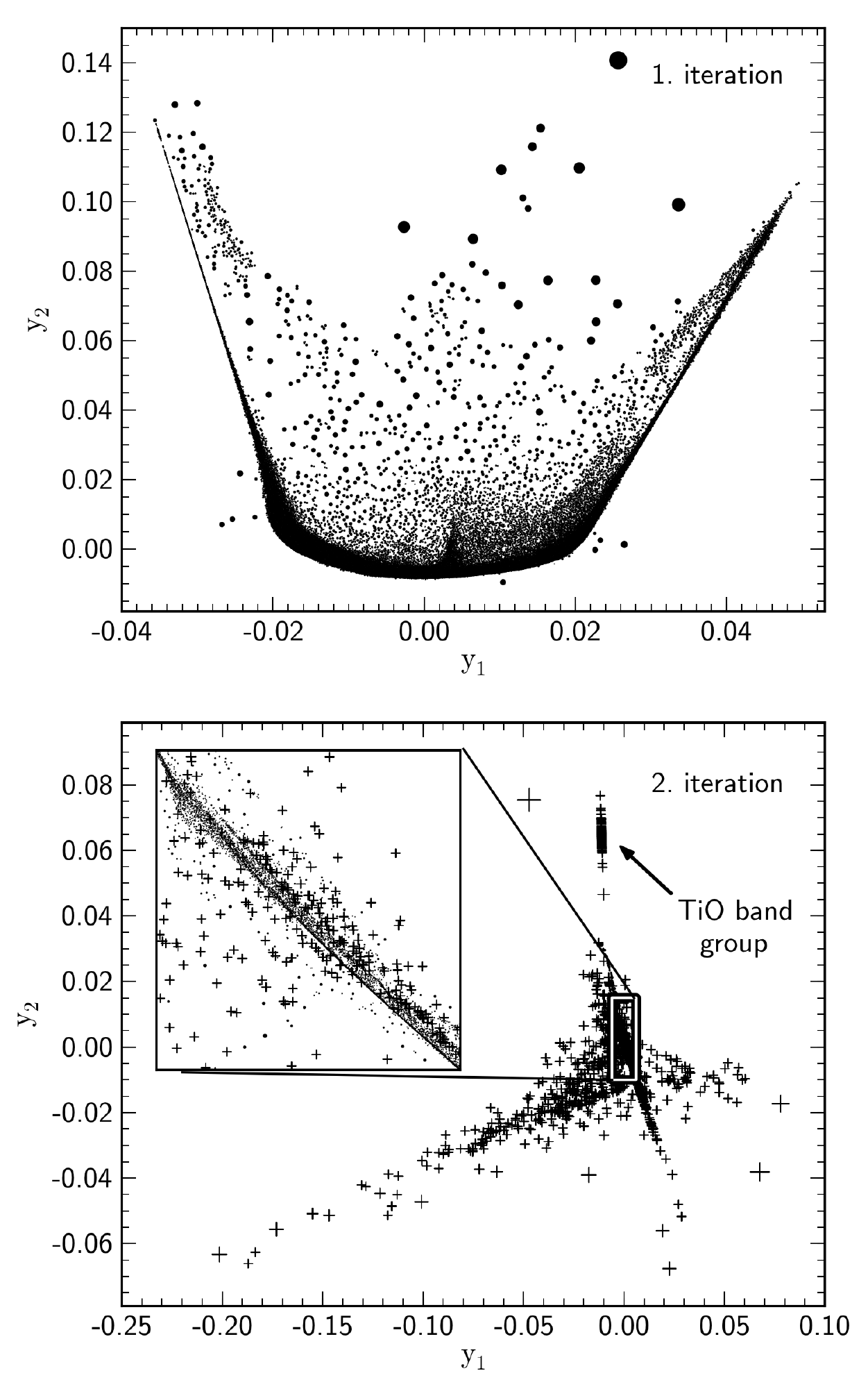}
\end{center}
\figcaption{First and second iteration of the classification procedure. The size of the symbols on both diagrams is scaled according to the distance to the nearest neighbor of each point for better viewing. For the sake of clarity not all points are shown. The inset in the bottom diagram shows the central region of the LLE projection map where it is possible to see the main arc around which the outliers are scattered. The dimensions on the axes are arbitrary.}
\label{fig:lle_iter}
\end{figure}

For this analysis we used the RAVE 101111 internal data release database that consists of 434,807 observations of 373,138 stars. Of all these spectra we analyzed only those with $\mathrm{S/N}>20$, and removed all spectra that are part of the first data release. The reason for the latter exclusion is that these spectra were polluted with second order light and are therefore not directly comparable to spectra recorded later when a blocking filter was installed. This selection lowered the number of spectra in our sample to 350,962. All spectra were treated individually and no connections between repeated observations of the same objects were taken into account during the analysis. 

\begin{figure}
\begin{center}
\includegraphics[width=0.5\textwidth]{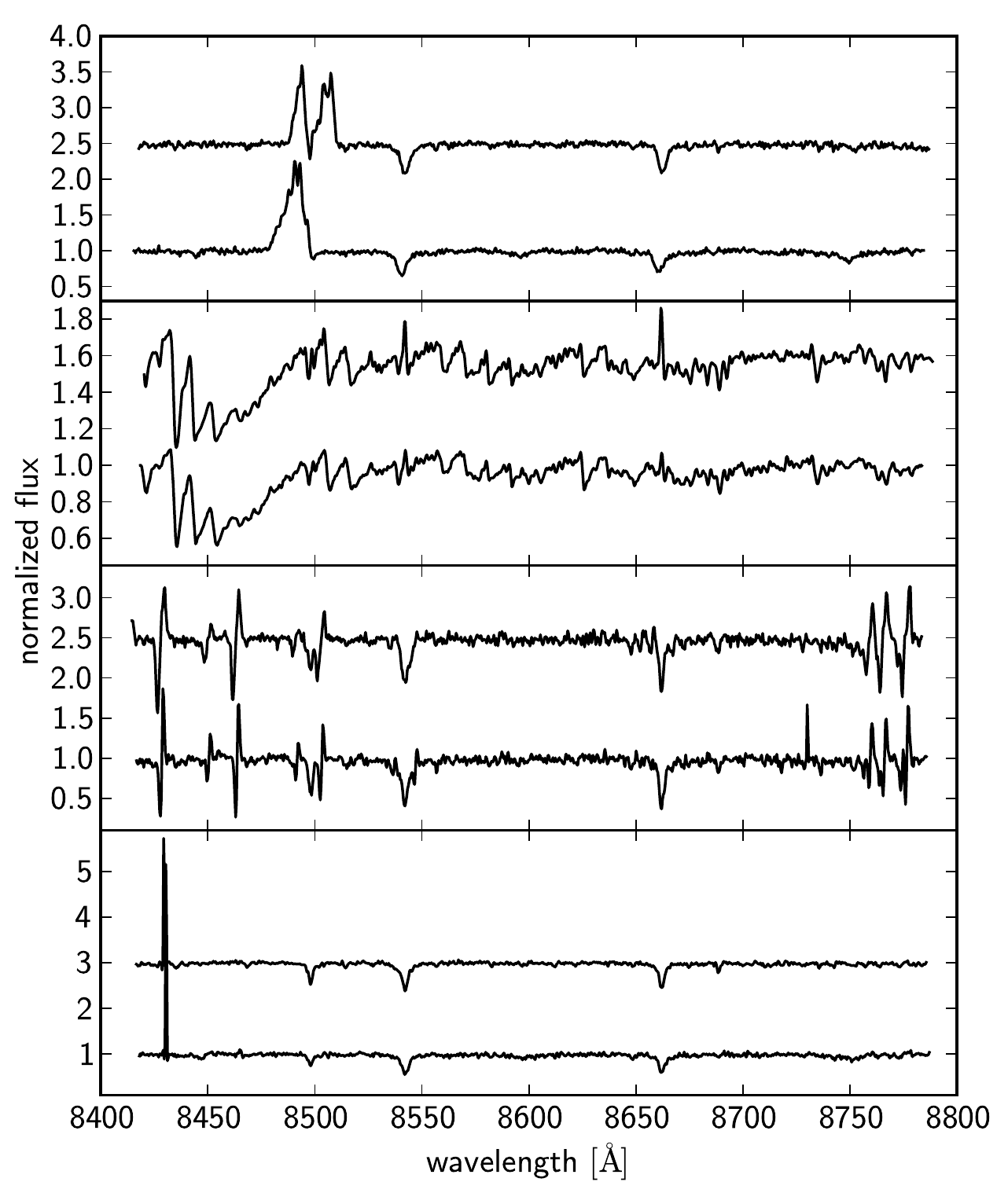}
\end{center}
\caption{Four distinct groups of outliers that were identified in the second iteration: spectra with a ghost signal, spectra with a deep $\mathrm{TiO}$ band, spectra with a characteristic oscillating signal and spectra with a strong spike in blue part of the spectral range.}
\label{fig:sec_iter_spec}
\end{figure}

Due to the arbitrary radial velocities of the stars, the RAVE spectra are sampled at different wavelength bins when they are transfered to the zero-velocity system. To make them comparable, we resampled them to the common wavelength range spanning from $8420\,\mathrm{\AA}$ to $8780\,\mathrm{\AA}$ with an equidistant step of $0.3\,\mathrm{\AA}$ using cubic splines. A few spectra did not cover the selected range, so we padded them with unity values on either side to cover the entire range. Slight oversampling (1200 wavelength bins instead of $\sim 1000$ bins RAVE spectra are usually sampled at) yields better results but the number of bins is on the other hand small enough not to hog the calculations. 

The first step towards producing a meaningful low dimensional projection of the whole sample is to define a well-sampled base subsample onto which all other spectra can be projected. Well-sampled refers to different spectral morphologies being equally represented in the subsample. The iterative process through which we generated the wanted base subsample was started with a random selection of 5000 spectra from the whole set of 350,962 spectra. This number was chosen to be large enough to include different types of spectra but still small enough so the calculation is performed quickly. Note that the random selection of spectra clearly violates the well-sampled assumption because different spectral morphologies are not represented equally, so this base subset was only used as a starting point. With the given subsample, we calculated the LLE projection onto a $d=3$ dimensional space, setting $k=20$ and $r=10^{-3}\mathrm{Tr}(\mathbf{C})$. The latter two values were chosen by trial and error and were shown to give the cleanest separation between the quasi-classes calculated with the cross-correlation method from \citet{2010AJ....140..184M}. Also, an example from \citet{2003SR} shows that the number of nearest neighbors is not critical and does not lead to significantly different projections as long as it is not too low or too high. Checking different cross-sections of the projection revealed that the projected manifold is mostly embedded in the first two dimensions (upper diagram of Figure~\ref{fig:lle_iter}) so in the following iterations we proceeded with projecting only onto $d=2$ dimensions.

To generate a better approximation of a well-sampled set, we sieved the first projection in such a way that densely populated areas in two dimensions were diluted, but sparsely populated regions containing spectra with rarer morphologies were not reduced. This was done in a separate iterative procedure. First, the two dimensional projection was overlaid with a randomly positioned uniform mesh and then a limited number of points from each mesh bin was advanced into the next iteration step. Continuing this process and varying the positions of the centers of mesh bins at each step effectively sieves the sample so that the distribution of the points is more and more uniform. This process was stopped when the number of remaining spectra fell just below 5000, yielding a base subsample for the next step of the main iteration. After that we repeated the previous step of projecting the remaining spectra from the full set onto the newly generated base. The projected map is completely different after the second iteration (Figure~\ref{fig:lle_iter}). Since the base subsample in the second iteration included a greater range of different spectra, the shape of the final projection in the second iteration was controlled by those points. Examining the underlying spectra of these extreme points (marked with $'+'$ in Figure~\ref{fig:lle_iter}) revealed that the majority of the spectra has a strong spike in the bluest part of wavelength range (bottom diagram of Figure~\ref{fig:sec_iter_spec}). Further investigation showed that this was clearly an observational or reduction error since it always plagued spectra recorded from the same few optical fibers. There are some additional morphological groups of spectra among the extreme outliers. One of them consists of spectra compromised by a specific reduction error and another of spectra having very deep $\mathrm{TiO}$ bands and largely offset radial velocity shifts (middle diagrams of Figure~\ref{fig:sec_iter_spec}). Note that the majority of $\mathrm{TiO}$ band spectra in the RAVE sample does not belong in this group. Among the outliers are also some spectra with faulty wavelength calibration and a strong ghost signal (top diagram of Figure~\ref{fig:sec_iter_spec}) that is caused by the light reflected from the detector, recollimated by the camera, reflected back by the grating and finally reimaged by the camera onto the detector \citep{2004S}. 
\begin{figure*}
\includegraphics[width=\textwidth]{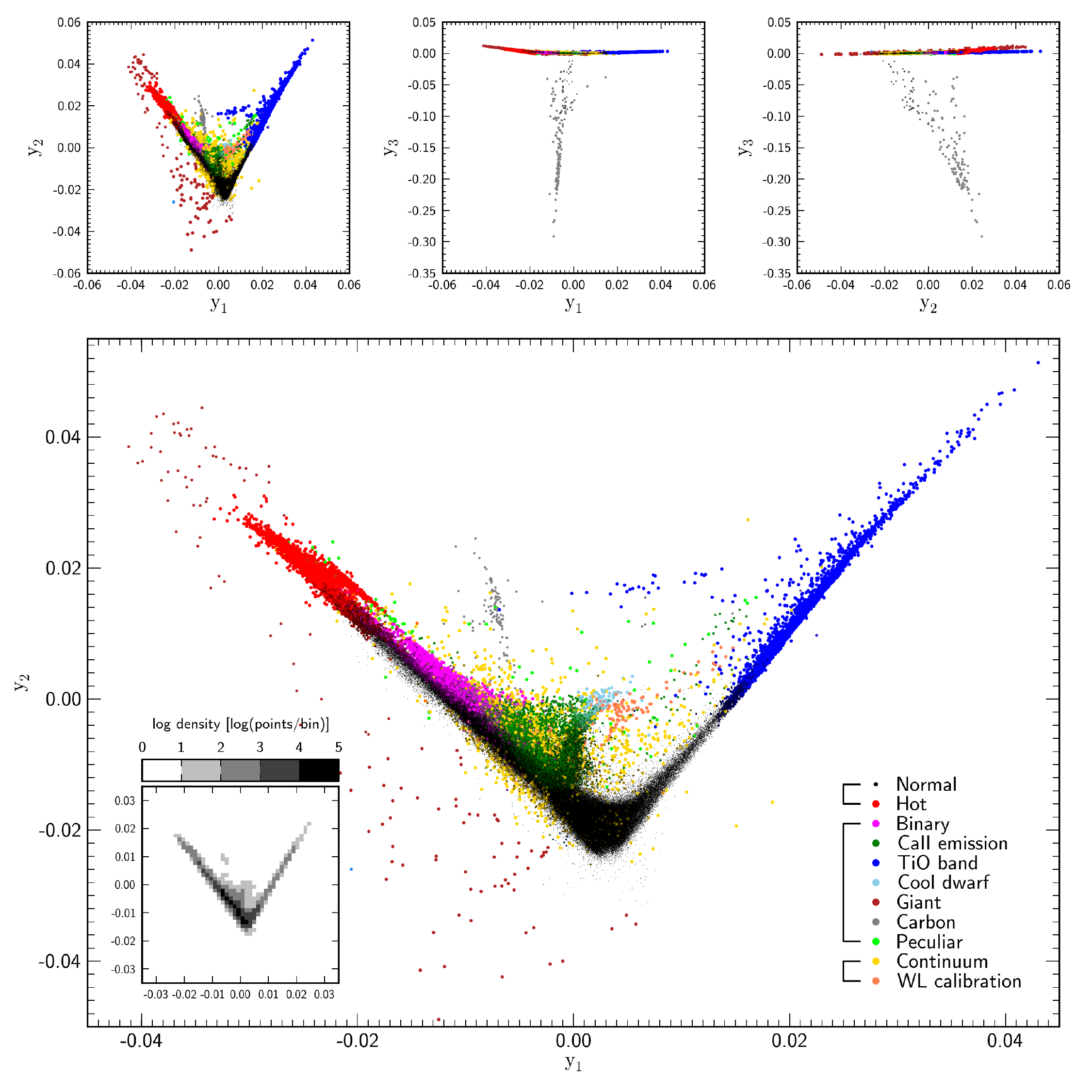}
\caption{Final LLE projection map. The top three diagrams show the cross-sections of the first three dimensions of the projected space, while the main diagram shows the higher resolution cross-section of the first two dimensions, where different regions can be best seen. Normal stars are intentionally represented with smaller symbols for the sake of clarity. Colors correspond to different spectral classes based on the first three largest weights. If the first three classification flags were different, the color was calculated as an weighted average of R(ed)G(reen)B(lue) values, where the weights were the same as returned by the LLE method. Note that this way of coloring is only used to show how the classes mix between each other. The bottom inset shows the $\log$ density of points on the main diagram.}
\label{fig:lle}
\end{figure*}

These spectra clearly drive the way the projection is rendered and overshadow the majority of the other spectra, so we decided to remove them from the sample. This was done by dropping the most extreme outliers after visually confirming that they had systematic problems or are from the $\mathrm{TiO}$ band group and to continue with the iteration process until the projected sample did not have any obvious outlying points anymore. A final projection that was produced as described is shown in Figure~\ref{fig:lle}.

For classifying the objects in the projected map, \citet{2011AJ....142..203D} suggested continuing the iterative process of picking out the outliers and grouping them together based on the different morphologies. Unfortunately, this approach only works for very distinct morphologies (Carbon stars in our case, see Sect.~\ref{sec:results}), but cannot coherently group classes that are connected continuously, i.e. are governed by some process that changes continuously. An example would be a double lined binary spectrum, which is easily identifiable if the Doppler separation between the pairs of spectral lines is greater than the resolution limit but such a spectrum can be easily mistaken for that of a single star if the opposite is true. Instead we proceeded with a different strategy. As before, we sieved the final $d=2$ projection until the overall number of spectra reached $\sim 5000$ and covered the first two dimensions as evenly as possible, creating a new base sample. Then we manually classified all spectra in this base sample by selecting different regions of the projection map from Figure~\ref{fig:lle} and assigning classification flags to spectra in each region. The classification was made easier by several studies of morphologies of normal and peculiar spectra already carried out for the \textit{Gaia} space mission \citep{1999BaltA...8...73M,1999A&AS..137..521M,2001A&A...378..477M, 2002ASPC..279...25M,2003ASPC..298..227M,2003ASPC..298..451P,2003ASPC..298..461R,
2004A&A...413..635M,2010AJ....140.1758T}. Since \textit{Gaia} shares a common spectral domain with the RAVE survey, the results of the studies can be directly applied to our sample. Spectra from the base sample were also compared to the existing solutions calculated by the RAVE pipeline for stellar parameters \citep{2011AJ....141..187S} to search for any signs of peculiarity. In the end, spectra were divided into 11 distinct classes discussed in the next section.

\begin{figure}
\begin{center}
\includegraphics[width=0.5\textwidth]{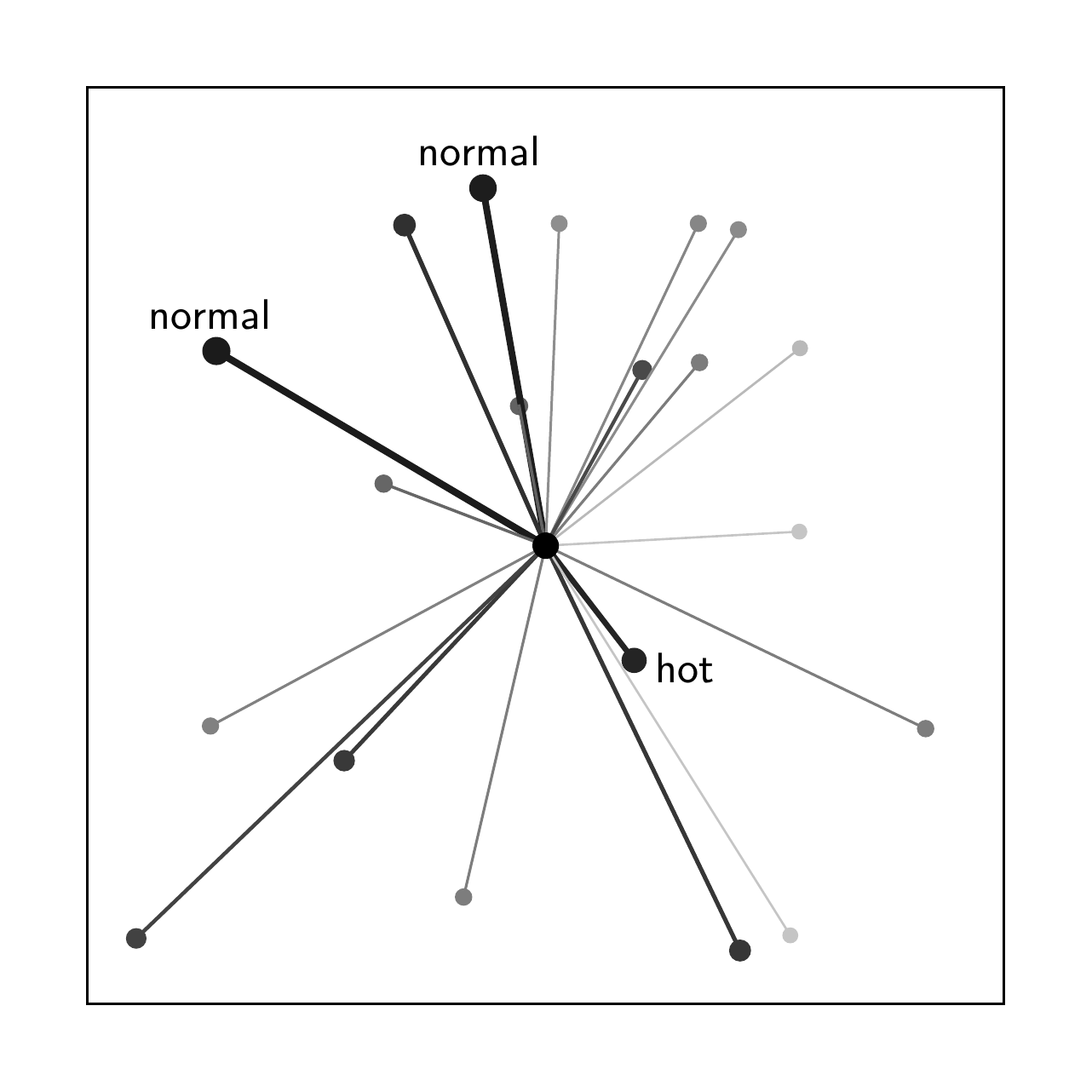}
\end{center}
\caption{Low-dimensional schematic representation of assigning classes. The central dot represents the projected spectrum being classified, while the other dots are its nearest neighbors from the two-dimensional base subset. Note that the dots are scattered randomly around the central spot in this representation but would otherwise be placed differently, depending on the position in the manifold shown in Figure~\ref{fig:lle}. The thickness of the connecting line, the darkness and the size of the symbol corresponds to the magnitude of the weight. In this example a classified spectrum has two normal star spectra and a hot star spectrum among its neighbors with the three largest weights, therefore it is considered to be on the border between the normal and hot classes ($y_1\approx -0.02$ and $y_2\approx 0.01$ in Figure~\ref{fig:lle}).}
\label{fig:classification}
\end{figure}

The automatic classification of the remaining spectra was conducted as follows. First, for each spectrum being classified, the $k=20$ nearest neighbors were found among the base sample and weights were calculated in the same way as described in Sect.~\ref{sec:lle}. Then the weights from the reconstruction of each spectrum were ordered according to their absolute values, from the largest to the smallest. Since the largest weights belong to those spectra in the base sample that are most similar to the spectrum being classified, it is possible to assign classes to new spectra based on the known classes of spectra from the base sample (Figure~\ref{fig:classification}). Of course, there is no guarantee that the reference base spectra of a given spectrum are from a single class, so multi-class classifications are possible. Examining the classification flags for a number of classified spectra revealed that in most cases only the first three largest weights and the corresponding base spectra are important. We decided to leave the final classification choice to the user. If only a single classification flag is preferred, there are multiple choices how to generate it. In our experience a good way of producing a single flag is to sum all weights corresponding to the same flag that are larger than half of the largest weight and to assign a final flag based on which of the summed flags has the highest score, i.e. the one with the largest sum of weights. The other option is to present the user with three flags corresponding to three largest weights and let the user decide on the final class.

\section{Morphological Classes of Spectra}\label{sec:results}

Based on the LLE projection we classified spectra into 11 distinct classes that can further be grouped into three supergroups. The first group represents the majority of the RAVE spectra and includes normal single stars. Normal in this case denotes all spectra for which it is possible to find a suitable counterpart in the library of synthetic spectra and for which one can therefore also calculate a reliable set of  parameters. We made a distinction between cooler and hotter stars where the separation between the two is based on the presence of Paschen series hydrogen lines which are normally found in spectra of stars with $T_\mathrm{eff}\ga 7000\,\mathrm{K}$. The second group consists of 8 different peculiar classes. The term peculiar in a broader sense marks all spectra that do not have a counterpart in the library of synthetic spectra and for which it is therefore not possible to infer their atmospheric parameters by simply finding the best matching synthetic spectrum. In this group there are different kinds of spectroscopic binary stars, stars with an observable emission component in the \ion{Ca}{2} lines that is a signature of chromospheric activity, cooler giant stars with a significant $\mathrm{TiO}$ molecular band\footnote{This group is equivalent to the one reviewed in Sect.~\ref{sec:proc}. Spectra in the excluded group have largely offset radial velocities and therefore look unique when compared to properly shifted $\mathrm{TiO}$ band spectra.}, hot and cool giants, cool normal stars $(T_\mathrm{eff}<3500\,\mathrm{K})$, carbon stars, and other types of peculiars. The last group consists of two classes of spectra with systematic errors. In the first class of this group are spectra with various continuum problems (oscillating continuum caused by poor continuum normalization, ghosting etc.). The second class contains spectra that were poorly calibrated in wavelength, but were not recognized as such in the initial iterations.

\subsection{Normal stars}

\begin{figure}
\begin{center}
\includegraphics[width=0.5\textwidth]{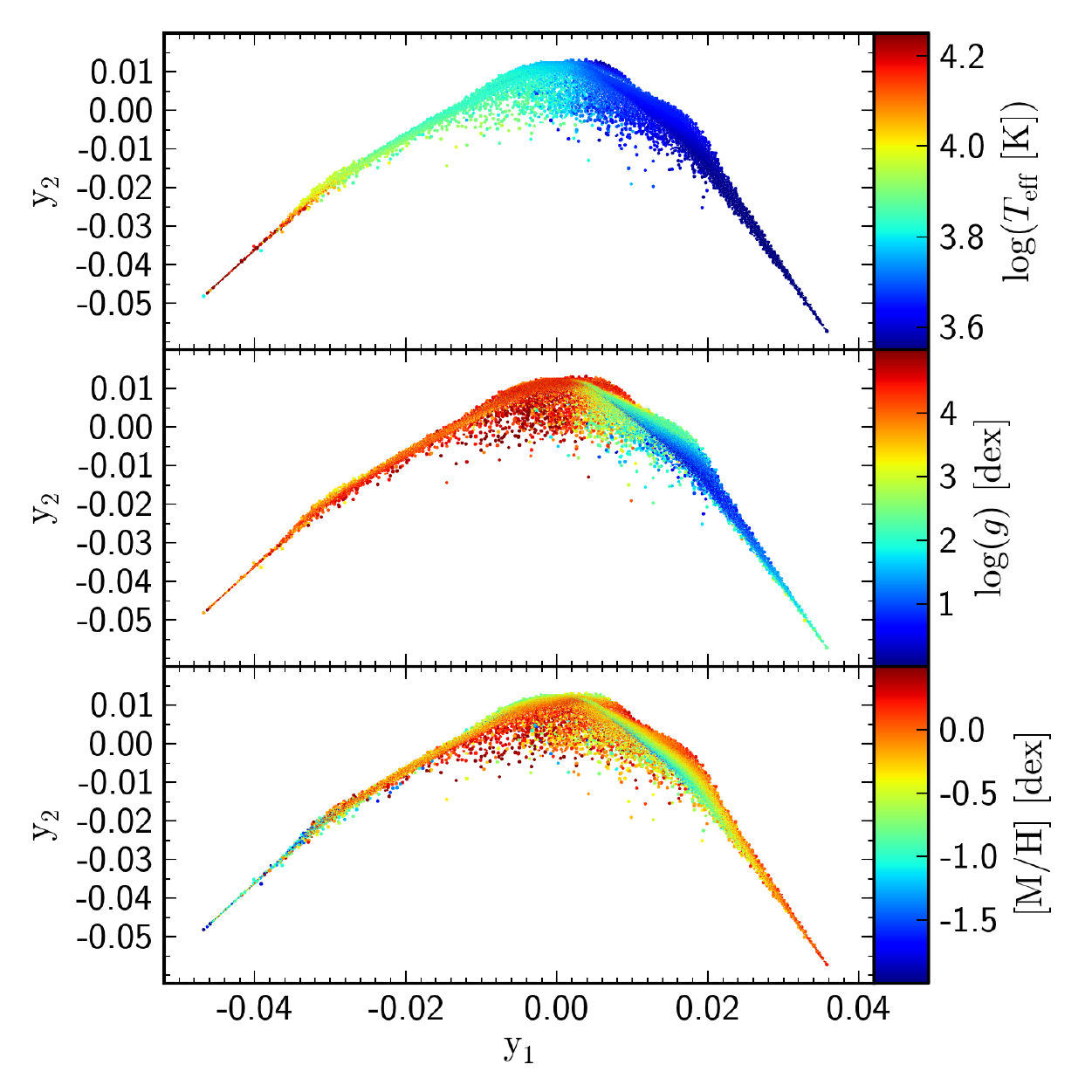}
\end{center}
\caption{Cross-section of the first two dimensions of LLE projection of the sample of normal single stars calculated with the $k=30$ nearest neighbors. The color scales in the diagrams represent different values of effective temperature $(T_\mathrm{eff})$, surface gravity $(\log(g))$ and metallicity $(\mathrm{[M/H]})$. The sizes of the points are scaled with respect to the density of the stars for clarity.}
\label{fig:lle_single}
\end{figure}
Single stars of various stellar types are the most common population in the RAVE sample. They contribute as much as $\sim 95\,\%$ of the spectra if their classification is based on the averaging method described above. Their large abundance in comparison to other morphological types can also be confirmed from the density plot shown in Figure~\ref{fig:lle}. As these are the spectra for which the atmospheric parameters are reliably estimated, we calculated the LLE projection for a sample of these stars alone to see if there are any correlations between the positions on the projection map and various parameters. The procedure is similar as before with the exception of the number of nearest neighbors that was set to $k=30$ since it produced slightly better results than $k=20$. Results are shown in Figure~\ref{fig:lle_single}.

From the top diagram it is evident that the leading parameter in the distribution of points in the two dimensional map is the effective temperature of the stars. The position along the arc can be used for a quick estimate of this parameter. A small number of points $(\sim 100)$ scattered below the main arc corresponds to misclassified spectra due to slight systematic problems like an oscillating continuum or the presence of spikes. Nevertheless, their number is negligible in comparison to number of the total spectra in this sample.

The distribution of points with respect to surface gravity is consistent with RAVE's bimodal population distribution. Roughly half of the whole sample is composed of dwarf stars and the other half of giants, a consequence of RAVE's magnitude limit \citep{2011AJ....141..187S}. The separation occurs just above $y_1=0$ where the values of $\log(g)$ become consistent with giant stars. The higher values of $\log(g)$ at the far right end of the arc are probably overestimated since $\mathrm{TiO}$ bands in spectra from this region become more pronounced, which complicates parameter estimation.

The diagram showing the metallicity of the stars is consistent with two populations as well. Particularly in the region that is populated by giants it is evident that the axis perpendicular to the main arc describes the metallicity. This effect is stronger with giants since their spectra show larger morphological changes with respect to metallicity than dwarfs, but it can also be observed in the part of the diagram populated by dwarfs. The diagram also highlights the underestimation of the metallicity of hot stars. This is a known problem and is attributed to the lack of strong metallic lines in these stars.

\subsection{Binary stars}

\begin{figure*}
\begin{center}
\includegraphics[width=0.95\textwidth]{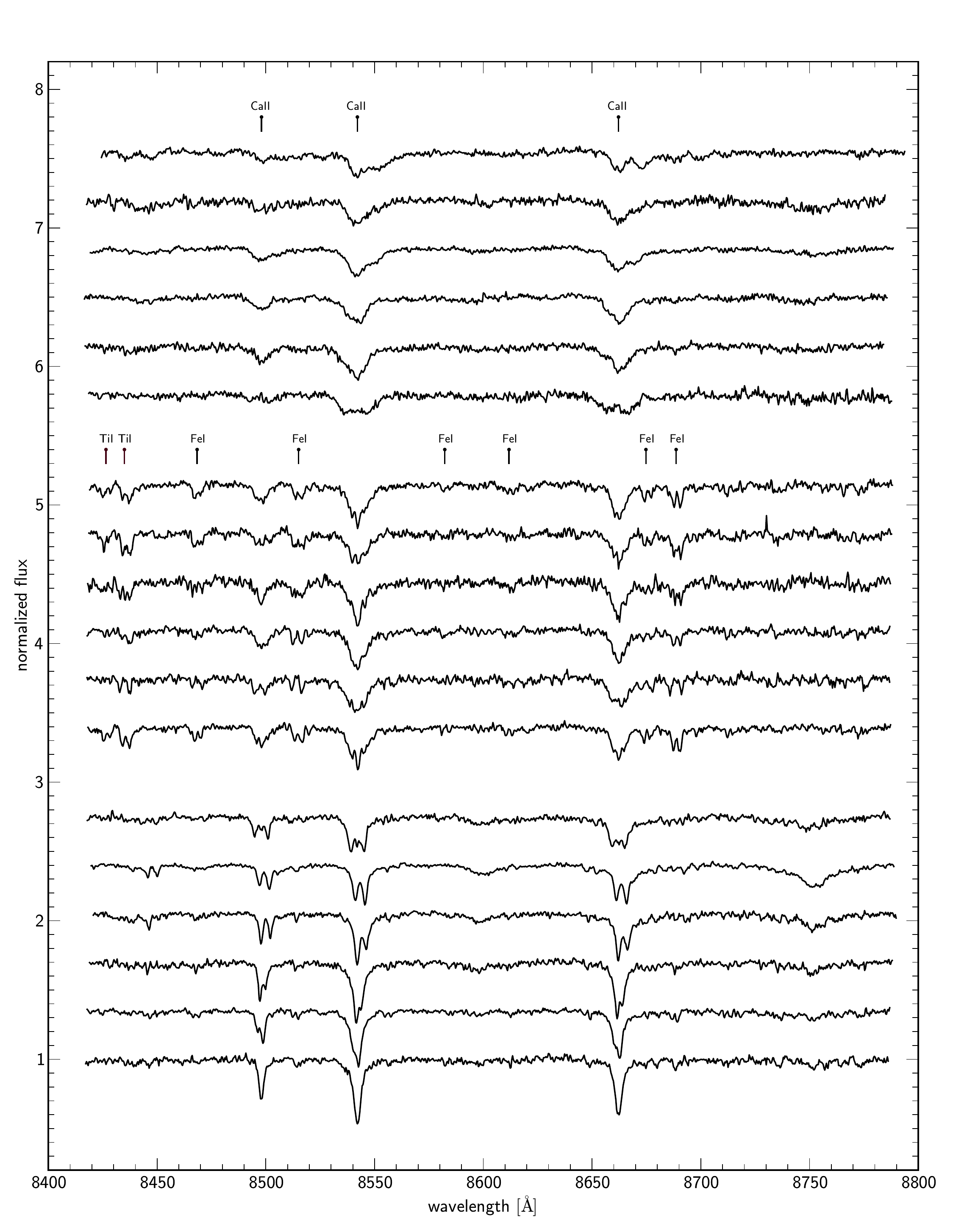}
\end{center}
\caption{A selection of of three most common types of binary star spectra. The bottom group shows five examples of regular SB2 spectra with a growing separation between the components. The sixth spectrum is a triple star where all three components can be observed spectroscopically. The middle group shows six examples of RS CVn type binary stars with active components. The top group depicts spectra of W UMa type binary stars. The marks identifying individual lines are centered at wavelengths measured in the rest system.}
\label{fig:bin}
\end{figure*}

Double-lined spectroscopic binary stars (SB2) represent $\la 1\,\%$ of RAVE population. The true number of such objects is hard to estimate due to their variable nature (i.e. if an SB2 is observed at a half phase, it spectrum looks very similar to a spectrum of a single star) and line-blending. However, the simulation from \citet{2010AJ....140..184M} found that the detection rate should be fairly high $(\sim 80\,\%)$ for systems with orbital periods shorter than $\approx 100\,\mathrm{days}$. In the LLE diagram they are close to hot metal-poor stars because their spectra share shallow spectral lines with those two classes. Morphologically, three different types are found among the binary sample (Figure~\ref{fig:bin}). Most common are the normal SB2s which are well represented by the sum of two Doppler shifted normal single star spectra. Another distinct group are binaries of the RS~CVn type where one or both of the components show chromospheric activity through emission in the \ion{Ca}{2} lines. In the third group are W~UMa type binary stars. Spectra of these short period binaries have only a few notable spectral lines that are broadened by corotation of the components with the system's period. 

\subsection{Chromospherically active stars / cool dwarfs}

\begin{figure*}
\includegraphics[width=\textwidth]{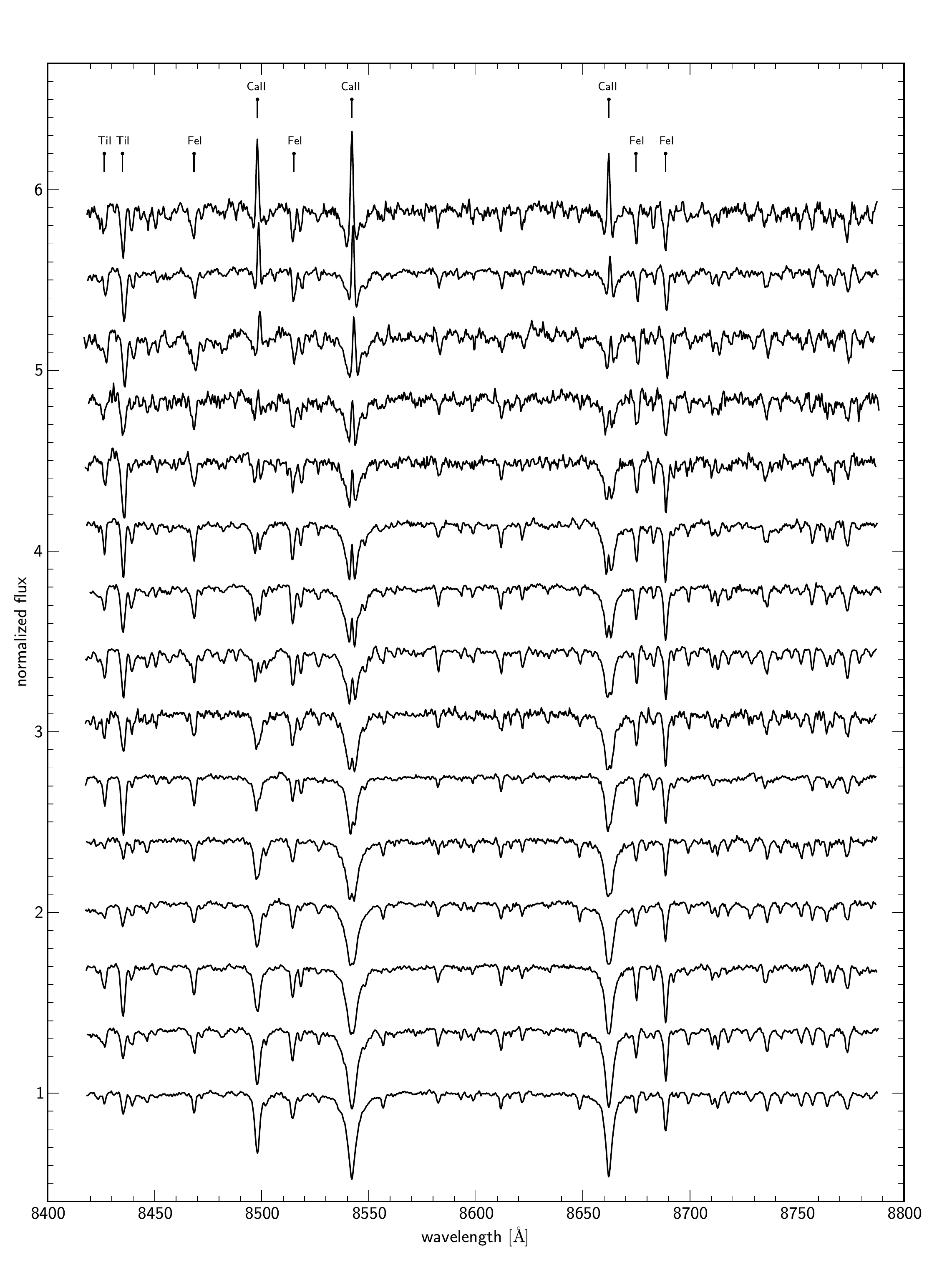}
\caption{A sequence of chromospherically active stars with growing amplitude of \ion{Ca}{2} emission (from bottom to top). The spectrum at the very bottom is only slightly different from what a common RAVE spectrum without chromospheric emission looks like.}
\label{fig:em}
\end{figure*}

Chromospherically active star are recognized by the emission component in the central part of the \ion{Ca}{2} triplet lines \citep{2005A&A...430..669A}. In the RAVE sample they are mostly dwarf stars of spectral type K but can also be hotter and cooler. The exact number of such stars in the RAVE sample is very difficult to assess due to their variability and the fact that in most stars this effect is small so only the most central parts of the \ion{Ca}{2} lines are slightly elevated. Nevertheless, if they are not identified and they are considered to be normal stars, this can lead to compromised values of atmospheric parameters. In comparison to other peculiar types they are the most abundant ones (between $2-3\,\%$ of all RAVE spectra). Their position on the LLE map is close to that of binaries since their \ion{Ca}{2} lines appear to be split. Thus their position overlaps with a branch of cool dwarf stars that are rare in the RAVE survey due to the magnitude selection cut. The level of chromospheric activity is a proxy for stellar ages \citep{2008ApJ...687.1264M} so these objects could be used as an independent age estimator. A sequence of several exemplary spectra with a growing level of chromospheric emission are shown in Figure~\ref{fig:em}.

Cool dwarfs on the other hand do not show any particularly interesting peculiarity  but were excluded from the main normal star set because their temperatures are lower $(T_\mathrm{eff}<3500\,\mathrm{K})$ and they are thus out of reach of the synthetic library used for modeling the spectra.

\subsection{TiO band stars}

\begin{figure*}
\includegraphics[width=\textwidth]{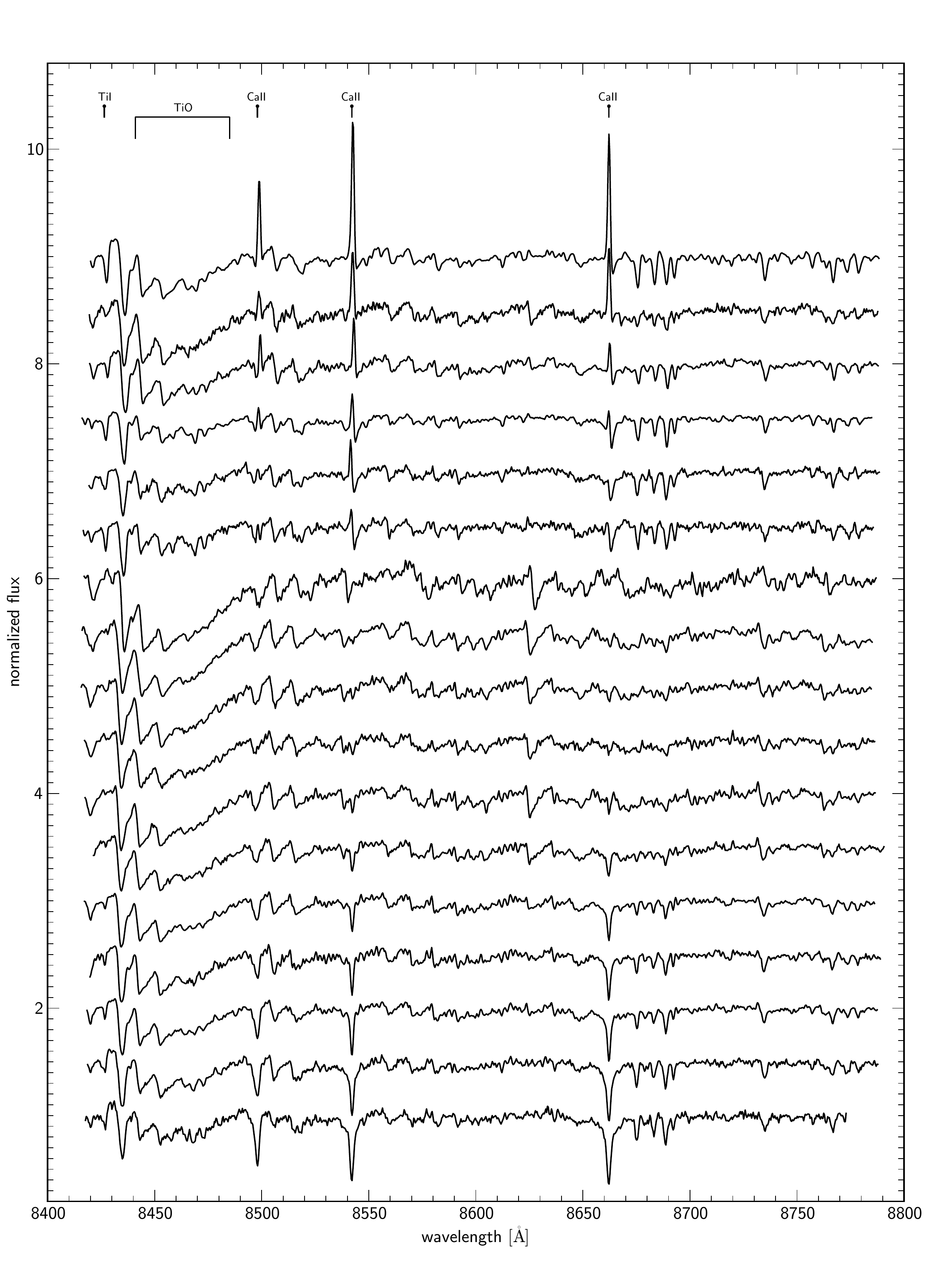}
\caption{A sequence of spectra with a strong $\mathrm{TiO}$ band. The top six spectra in addition to the molecular band also have \ion{Ca}{2} emission.}
\label{fig:tio}
\end{figure*}

Cool giants stars with a deep $\mathrm{TiO}$ molecular band in the bluest part of the spectrum are relatively abundant in the RAVE sample $(\sim 1\,\%)$. Some of the stars in this class are non-peculiar giants and are excluded (same as cool dwarfs) due to the lack of synthetic spectra in the library with which they could be modeled. According to the SIMBAD database, there are also many known Mira type pulsating variables in this group. A sequence of selected spectra of this type is shown in Figure~\ref{fig:tio}. Starting from the bottom, the leading parameter that changes the shape of the spectra of the first group is most likely a decreasing effective temperature. The top six spectra in the second group exhibit an emission component in the \ion{Ca}{2} lines in addition to the $\mathrm{TiO}$ band.

\subsection{Peculiar giants}

\begin{figure*}
\includegraphics[width=\textwidth]{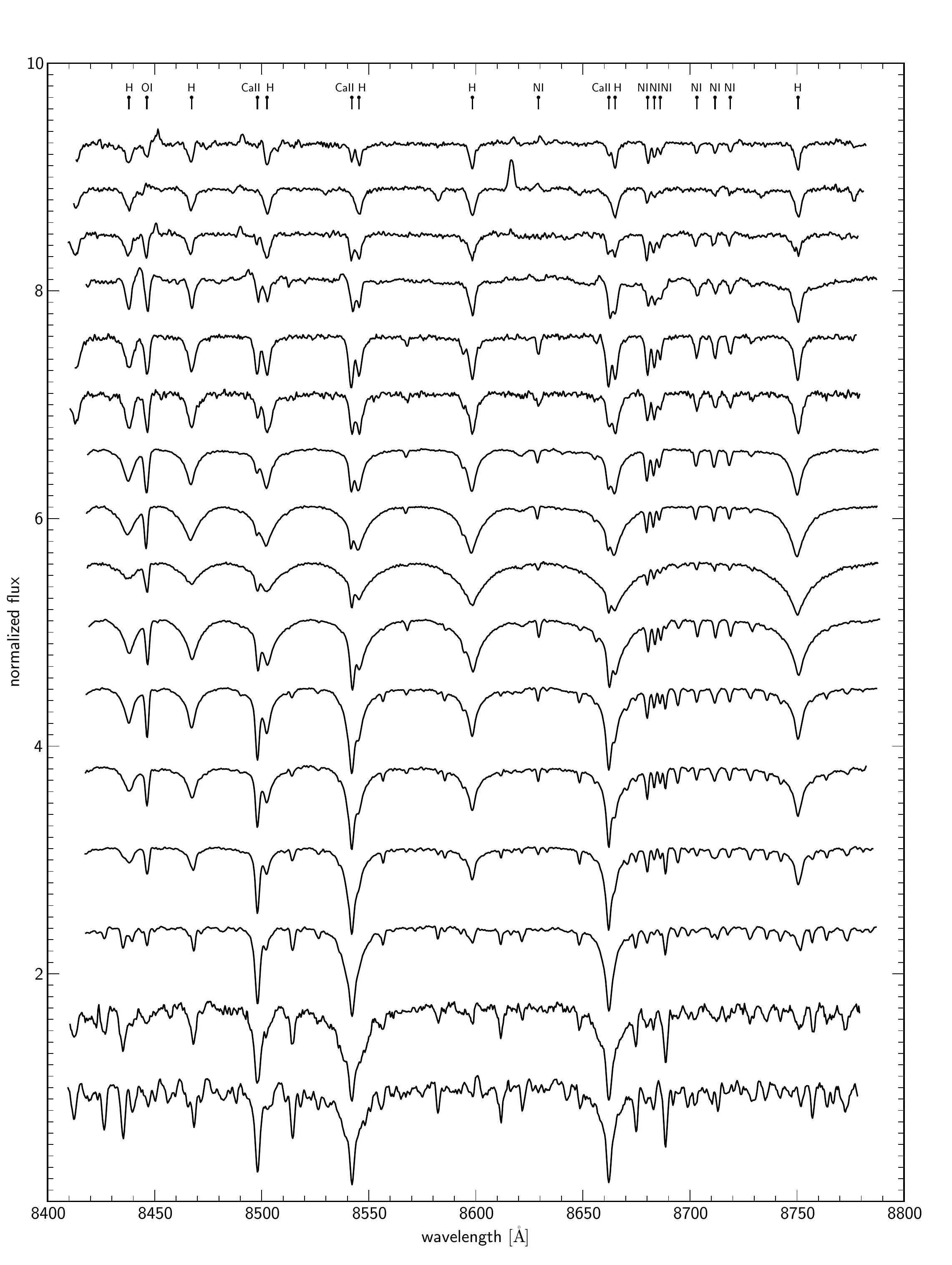}
\caption{A sequence of spectra of peculiar giant stars.}
\label{fig:giant}
\end{figure*}

In this case we use the label giant for the separated spectra that lie in the bottom left corner of the projection map in Figure~\ref{fig:lle}. Again, they are not all necessarily peculiar in the classical sense but their morphologies are very rare in the RAVE sample which is why they are projected separately from the rest of the normal stars. Also, the best matching spectra from the synthetic library as predicted by the parameter estimation pipeline might not sufficiently describe these spectra, which is particular problematic given their usually high $\mathrm{S/N}$ ratio, so  
their atmospheric parameters are likely unreliable.. Particularly interesting are the hotter examples (top spectra in Figure~\ref{fig:giant}) with deep \ion{N}{1} and narrow Paschen hydrogen lines. Overall they present a negligible contribution to the whole sample with $\sim 100$ objects total.

\subsection{Carbon stars}

\begin{figure*}
\includegraphics[width=\textwidth]{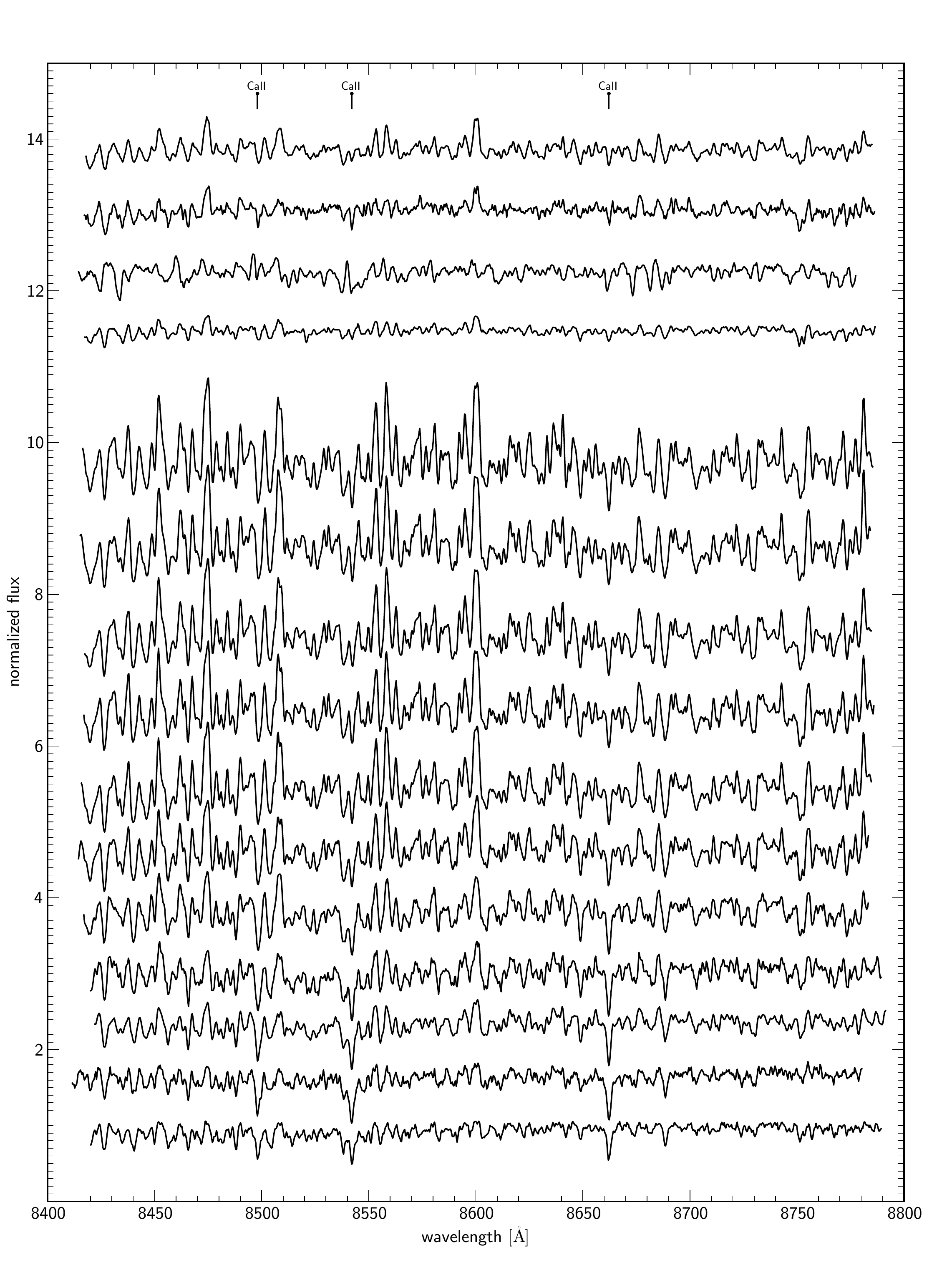}
\caption{A sequence of spectra of carbon stars.}
\label{fig:carbon}
\end{figure*}

Carbon stars are another very distinct but rare morphological group in the RAVE sample. Due to the presence of many CN lines the morphology of the spectra of these stars is very different from other types. Consequently, their position on the LLE map is isolated and they are also the only group that has a strong third component (Figure~\ref{fig:lle}). Altogether, around 100 such stars are present in the database. A sequence of different spectra of already known carbon stars (according to SIMBAD database) is shown in Figure~\ref{fig:carbon}.

\subsection{Other peculiar stars}

\begin{figure*}
\includegraphics[width=\textwidth]{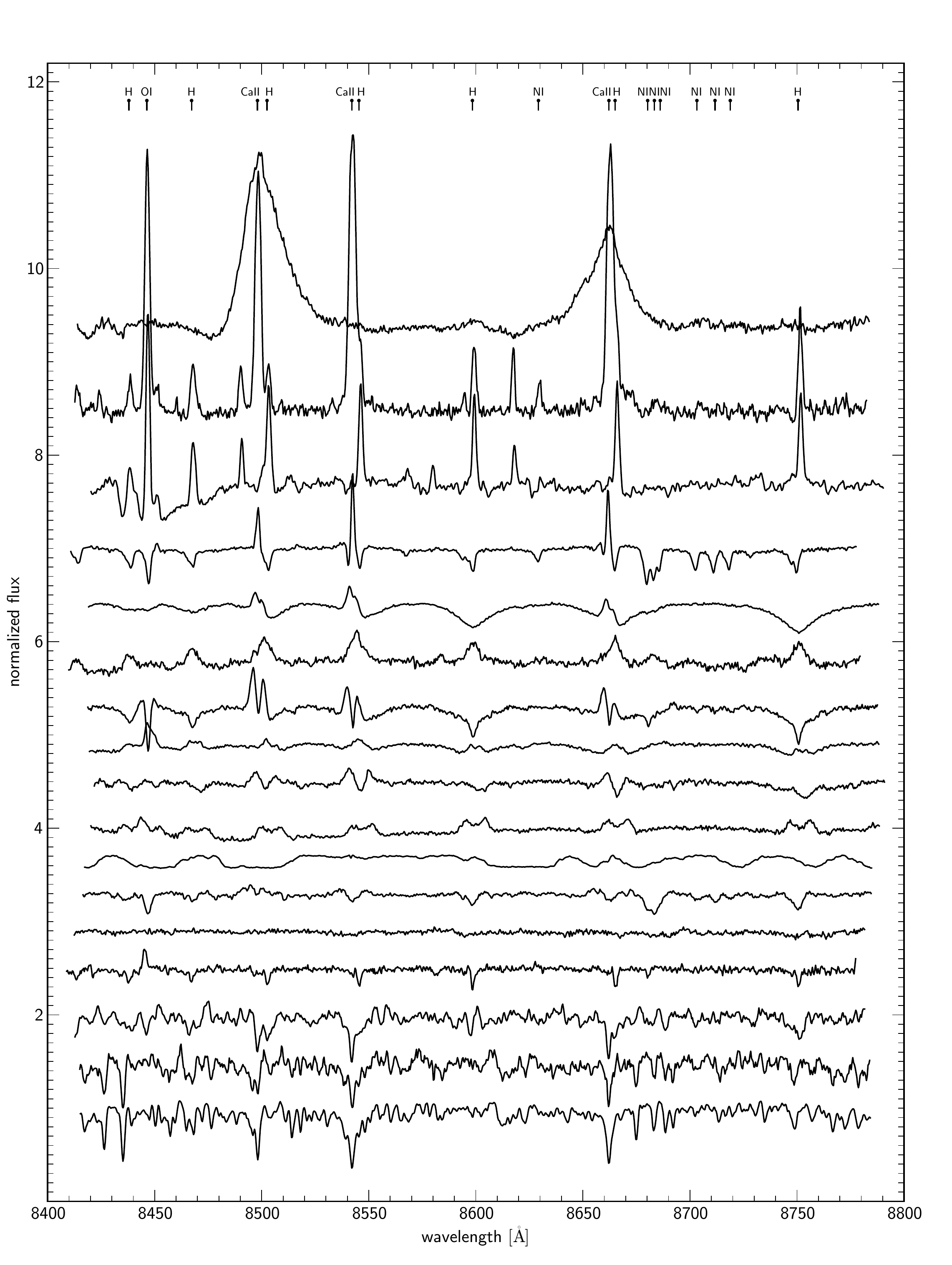}
\caption{A sequence of spectra of other types (mostly unidentified) of peculiar stars.}
\label{fig:pec}
\end{figure*}

This class includes the rest of the peculiar spectra. Due to their diverse morphologies and sparseness it is not possible to group and order them coherently. Some of the spectra from this class have a strong emission component in either \ion{Ca}{2} or hydrogen lines. The list includes a couple of known Wolf-Rayet stars, Be stars, and other types of variables, but there are no more than a few hundred such objects in the classified sample. A selection of higher $\mathrm{S/N}$ spectra from this class is shown in Figure~\ref{fig:pec}. Due to the limited amount of information available from a single RAVE spectrum we were unable to unambiguously identified most of previously unknown peculiars from this group.

\subsection{Problematic spectra}

Spectra with various artifacts (oscillating continuum, ghosts, spikes etc.) or wrong wavelength calibration that certainly influence the fitting procedure and hence their atmospheric parameters and radial velocities account for less than $1\,\%$ of all RAVE spectra. Their position on the LLE map is scattered around the classes, where  
the spectra that have no problems reside.

\section{Summary and Conclusions}\label{sec:sum}

The results of this study have shown that LLE can be used as an efficient classification tool. It is able to project the complex morphology of the RAVE stellar spectra onto a two dimensional space and to preserve the relations between neighboring points. 

The classification procedure was made in several steps by first repeatedly projecting the selected sample onto a low dimensional space and culling of the most extreme outliers, for which it turned out that are mostly spectra with characteristic systematic problems. Culling is a necessary step since the projection heavily depends on the input sample and leaving the outliers in the sample overshadows the more interesting morphologies of other spectra. When the final classification map was calculated, a base sample of  $\sim 5000$ of the most representative spectra were selected by a sieving  process that diluted the densely populated area of the two dimensional map, but left the sparsely populated areas untouched. After this, the sub-selection was manually classified by assigning classification flags from 11 distinct morphological classes to all spectra in the subsample. Classification flags of the rest of the $\sim 345,000$ spectra were set by finding the nearest neighbors from the base set and relating the classification flags of the neighbors with the highest weights to the final class of each spectrum. Since there is no unique way of how flags get assigned, the final choice of classification is left to the user. There are two possibilities: either use a single averaged flag or rely on the first three flags with the highest corresponding weights. Both ways have their uses. The first one is somewhat biased toward normal stars (the majority class) and so some slightly peculiar spectra, i.e. chromospherically active stars with only a minor \ion{Ca}{2} emission component might be flagged as normal. Since the effect of the emission component in this example is small, radial velocities and atmospheric parameters can be trusted relatively well. On the other hand, in this example there will likely be at least one active star flag among the top three so it can be identified and further investigated by a user interested in such peculiars. Either way, the most morphologically different spectra belonging to peculiar stars or stars with spectra with systematic problems are flagged appropriately, leaving around $\sim 90-95\,\%$ of spectra in the normal star class.

The analysis and projection of an isolated sample of normal single stars (selected according to the first criterion) shows that the shape of the manifold in two dimension correlates with all three major parameters: $T_\mathrm{eff}$, $\log(g)$ and $\mathrm{[M/H]}$. It also separates between two distinct populations in the RAVE sample (giants and dwarfs) relatively well. The amount of obvious outliers that are highlighted by the projection map is small (several 100), confirming the accuracy of the classification.

Around $5-10\,\%$ of the spectra turned out to belong to peculiar stars where the label peculiar denotes all spectra that are plausible (i.e. not problematic) but do not conform to the single normal star class. There are three major groups that comprise the majority of all peculiars: double-lined spectroscopic binaries, chromospherically active stars  and cool stars with significant $\mathrm{TiO}$ bands. In addition, there are a few minor groups like carbon stars, some rare giants, cool dwarfs and some other peculiars. The first three groups offer the potential for further research due to their abundance and there is already an observational program under way to explore the most interesting spectra in more detail and also at other wavelengths not covered by RAVE. Particularly, the sample of chromospherically active stars seems interesting due to the link between the level of chromospheric activity and stellar ages which can be exploited for age estimation. We plan to treat the spectra from peculiar groups in more detail in the series of forthcoming papers and add the classification flags to the future data releases.

All in all, we have demonstrated that LLE is a very appealing method when dealing with stellar or other types of astronomical spectra. With its relatively simple machinery it is able to represent the complex morphological properties of spectra in a very low dimension space, giving an opportunity for efficient discoveries of hidden features. We have also shown that this method can be used efficiently for consistent classification purposes without having to rely on external information.

\acknowledgements
Funding for RAVE has been provided by: the Australian Astronomical
Observatory; the Leibniz-Institut fuer Astrophysik Potsdam (AIP); the
Australian National University; the Australian Research Council; the
French National Research Agency; the German Research Foundation (SPP 
1177 and SFB 881); the European Research Council (ERC-StG 240271 
Galactica); the Istituto Nazionale di Astrofisica at Padova; The Johns 
Hopkins University; the National Science Foundation of the USA 
(AST-0908326); the W. M. Keck foundation; the Macquarie University; the 
Netherlands Research School for Astronomy; the Natural Sciences and 
Engineering Research Council of Canada; the Slovenian Research Agency; 
the Swiss National Science Foundation; the Science \& Technology 
Facilities Council of the UK; Opticon; Strasbourg Observatory; and the 
Universities of Groningen, Heidelberg and Sydney. The RAVE web site is 
at http://www.rave-survey.org.


\begin{thebibliography}{}
\bibitem[Abazajian et al.(2009)]{2009ApJS..182..543A} {Abazajian}, K.~N., {Adelman-McCarthy}, J.~K., {Ag{\"u}eros}, M.~A. et al. 2009, \apjs, 182, 543
\bibitem[Andretta et al.(2005)]{2005A&A...430..669A} {Andretta}, V., {Bus{\`a}}, I., {Gomez}, M.~T. \& {Terranegra}, L. 2005, \aap, 430, 669
\bibitem[Bailer-Jones et al.(1998)]{1998MNRAS.298..361B} {Bailer-Jones}, C.~A.~L., {Irwin}, M. \& {von Hippel}, T. 1998, \mnras, 298, 361
\bibitem[Colless et al.(2001)]{2001MNRAS.328.1039C} {Colless}, M., {Dalton}, G., {Maddox}, S. et al. 2001, \mnras, 328, 1039
\bibitem[Connolly et al.(1995)]{1995AJ....110.1071C} {Connolly}, A.~J., {Szalay}, A.~S., {Bershady}, M.~A., {Kinney}, A.~L. \& {Calzetti}, D. 1995, \aj, 110, 1071
\bibitem[Daniel et al.(2011)]{2011AJ....142..203D} {Daniel}, S.~F., {Connolly}, A., {Schneider}, J., {Vanderplas}, J. \& {Xiong}, L. 2011, \aj, 142, 203
\bibitem[de Ridder \& Duin(2002)]{2002dRD} {de Ridder}, S. \& Duin, R. 2002, Pattern Recognition Group, Department of Science and Technology, Delft University of Technology, Technical Report RH-2002-01
\bibitem[Gulati et al.(1994)]{1994ApJ...426..340G} {Gulati}, R.~K., {Gupta}, R., {Gothoskar}, P. \& {Khobragade}, S. 1994, \apj, 426, 340
\bibitem[Ibata \& Irwin(1997)]{1997AJ....113.1865I} {Ibata}, R.~A. \& {Irwin}, M.~J. 1997, \aj, 113, 1865
\bibitem[Jones et al.(2004)]{2004MNRAS.355..747J} {Jones}, D.~H., {Saunders}, W., {Colless}, M. et al. 2004, \mnras, 355, 747
\bibitem[Mamajek \& Hillenbrand(2008)]{2008ApJ...687.1264M} {Mamajek}, E.~E. \& {Hillenbrand}, L.~A. 2008, \apj, 687, 1264
\bibitem[Marrese et al.(2004)]{2004A&A...413..635M} {Marrese}, P.~M., {Munari}, U., {Siviero}, A. et al. 2004, \mnras, 413, 635
\bibitem[Matijevi{\v c} et al.(2010)]{2010AJ....140..184M} {Matijevi{\v c}}, G., {Zwitter}, T., {Munari}, U. et al. 2010, \aj, 140, 184
\bibitem[McGurk et al.(2010)]{2010AJ....139.1261M} {McGurk}, R.~C., {Kimball}, A.~E. \& {Ivezi{\'c}}, {\v Z} 2010, \aj, 139, 1261
\bibitem[Muja \& Lowe(2009)]{muja_flann_2009} {Muja}, M. \& {Lowe}, D.~G. 2009, in proceedings of International Conference on Computer Vision Theory and Application VISSAPP'09, 331
\bibitem[Munari(1999)]{1999BaltA...8...73M} {Munari}, U. 1999, {Baltic Astronomy}, 8, 73
\bibitem[Munari \& Tomasella(1999)]{1999A&AS..137..521M} {Munari}, U. \& {Tomasella}, L. 1999, \aaps, 137, 521
\bibitem[Munari et al.(2001)]{2001A&A...378..477M} {Munari}, U., {Tomov}, T., {Zwitter}, T. et al. 2001, \aap, 378, 477
\bibitem[Munari(2002)]{2002ASPC..279...25M} {Munari}, U. 2002, in Far-Red Spectroscopy of Peculiar Stars and the GAIA Mission, ed. {C.~A.~Tout \& W.~van Hamme}, ASP Conf. Ser. 279, 25
\bibitem[Munari(2003)]{2003ASPC..298..227M} {Munari}, U. 2003, in GAIA spectroscopy of peculiar and variable stars, ed. {Munari}, U., ASP Conf. Ser. 298, 227
\bibitem[Munari et al.(2005)]{2005A&A...442.1127M} Munari, U., Sordo, R., Castelli, F. \& Zwitter, T. 2005, \aap, 442, 1127
\bibitem[Munari et al.(2009)]{2009A&A...503..511M} {Munari}, U. and {Siviero}, A. and {Bienaym{\'e}}, O. et al. 2009, \aap, 503, 511
\bibitem[Pavlenko et al.(2003)]{2003ASPC..298..451P} {Pavlenko}, Y.~V., {Marrese}, P.~M. \& {Munari}, U. 2003, in GAIA Spectroscopy: Science and Technology, ed. {Munari}, U., ASP Conf. Ser. 298, 451
\bibitem[Ragaini et al.(2003)]{2003ASPC..298..461R} {Ragaini}, S. and {Andretta}, V. and {Gomez}, M.~T. et al. 2003, in GAIA Spectroscopy: Science and Technology, ed. {Munari}, U., ASP Conf. Ser. 298, 461
\bibitem[Roweis \& Saul(2000)]{2000Sci...290.2323R} {Roweis}, S.~ T. \& {Saul}, L.~K. 2000, Science, 290, 2323
\bibitem[Saul \& Roweis(2003)]{2003SR} {Saul}, L.~K. \& {Roweis}, S.~ T. 2003, Journal of Machine Learning Research, 4, 119
\bibitem[Saunders et al.(2004)]{2004S} {Saunders}, W., {Bridges}, T., {Gillingham}, P. et al. 2004, AAOmega: a scientific and optical overview, doi:10.1.1.150.7232
\bibitem[Siebert et al.(2011)]{2011AJ....141..187S} {Siebert}, A., {Williams}, M.~E.~K., {Siviero}, A. et al. 2011, \aj, 141, 187
\bibitem[Steinmetz et al.(2006)]{2006AJ....132.1645S} {Steinmetz}, M., {Zwitter}, T., {Siebert}, A. et al. 2006, \aj, 132, 1645
\bibitem[Tomasella et al.(2010)]{2010AJ....140.1758T} {Tomasella}, L., {Munari}, U. \& {Zwitter}, T. et al. 2010, \aj, 140, 1758
\bibitem[VanderPlas \& Connolly (2009)]{2009AJ....138.1365V} {VanderPlas}, J.~T. \& {Connolly}, A.~J. 2009, \aj, 138, 1365
\bibitem[von Hippel et al.(1994)]{1994MNRAS.269...97V} {von Hippel}, T., {Storrie-Lombardi}, L.~J., {Storrie-Lombardi}, M.~C. \& 	{Irwin}, M.~J. 1994, \mnras, 269, 97
\end{thebibliography}
\end{document}